%% file: novasci.tex
\documentclass[11pt,twoside,letterpaper]{article}

\usepackage{graphicx}
\usepackage{subfigure}
\usepackage[round,sort&compress]{natbib}
\usepackage{authblk}
\usepackage[nolists,nomarkers]{endfloat}

\graphicspath{{figs/}}

\include{novascimacros}
\include{xrefmacros}

\usepackage{times,fancyhdr}

\title{Noise and Neuronal Heterogeneity}

\author{Michael J. Barber
	\thanks{Present address: 
				ARC systems research GmbH, 
				Donau-City-Stra{\ss}e 1,
				1220 Vienna,
				Austria}}
\affil{Universidade da Madeira,
		Centro de Ci\^encias Matem\'aticas,
		Campus Universit\'ario da Penteada,
		9000-390 Funchal,
		Portugal,
		\email{michael.barber@arcs.ac.at}}
\author{Manfred L. Ristig}
\affil{Institut f\"{u}r Theoretische Physik, 
	Universit\"{a}t zu K\"{o}ln,
	D-50937 K\"{o}ln,
	Germany,
	\email{ristig@thp.uni-koeln.de]}}

\date{}

\begin{document}

\pagestyle{fancy}
\fancyhead{} 
\fancyhead[EC]{M. J. Barber and M. L. Ristig}
\fancyhead[EL,OR]{\thepage}
\fancyhead[OC]{Noise and Neuronal Heterogeneity}
\fancyfoot{} 
\renewcommand\headrulewidth{0.5pt}
\addtolength{\headheight}{2pt} 

\maketitle

\begin{abstract}

	We consider signal transaction in a simple neuronal model featuring 
	intrinsic noise. The presence of noise limits the precision of neural 
	responses and impacts the quality of neural signal transduction. We 
	assess the signal transduction quality in relation to the level of 
	noise, and show it to be maximized by a non-zero level of noise, 
	analogous to the stochastic resonance effect. The quality enhancement 
	occurs for a finite range of stimuli to a single neuron; we show 
	how to construct networks of neurons that extend the range. The range 
	increases more rapidly with network size when we make use of 
	heterogeneous populations of neurons with a variety of thresholds, 
	rather than homogeneous populations of neurons all with the same 
	threshold. The limited precision of neural responses thus can have a 
	direct effect on the optimal network structure, with diverse 
	functional properties of the constituent neurons supporting an 
	economical information processing strategy that reduces the metabolic 
	costs of handling a broad class of stimuli.

\end{abstract}

\section{Introduction}\label{sec:intro}

Neural network models are often constructed of simple units.  
Typically,  model neurons have a particular threshold or bias, and 
saturate to a fixed value for either strong or weak inputs.  
Some such models can in fact be derived by systematic approximations of 
more detailed models such as the Hodgkin-Huxley model \citep{AbbKep:1990}; 
many other models are derived from alternative heuristic or phenomenological 
assumptions. Networks of even the simplest models are well known to be capable of 
representing complex functions.

In this chapter, we investigate the 
degree to which the simple dynamics of  an individual 
unit limit the 
inputs that can be processed, and how these 
limitations can be 
overcome.
To do this, we  consider the response of model neurons to 
a variety
of stimuli.  The precision of the responses for biological neurons
have been shown to be rather limited, typically in the range of a few bits
for each action potential or ``spike'' \citep{RieWarRuyBia:1997}.
We mimic this limited precision in the model neurons by including noise in the 
systems.  We focus on intrinsic neuronal noise which has identical statistics 
for each of the units in the neural network.

Noise is usually viewed 
as limiting the sensitivity of a system, but 
nonlinear systems can 
react to noise in surprising ways.
Perhaps the best known of these is the phenomenon known as stochastic resonance (SR),
wherein
an optimal response to weak or subthreshold signals is observed when a 
non-zero 
level of noise is added to the system \citep{GamHanJunMar:1998}.  
For example, a noise-free, 
subthreshold neuronal input can occasionally become 
suprathreshold when noise 
is added, allowing some character of the 
input signal to be detected. 
SR has been observed and investigated in many systems, 
ranging from resonant cavities to neural 
networks to the onset of 
ice ages 
\citep[see, \eg,][]{MosPei:1995,WieMos:1995,BezVod:1997,GaiNeiColMos:1997,%
GoyHan:2000,JunShu:2001,SchGoyHan:2001,WenObe:2003}.

\citet{ColChoImh:1995a} showed, in a 
summing network 
of identical Fitzhugh-Nagumo model neurons, that 
an 
emergent property of SR in multi-component systems 
is that the 
enhancement of the response becomes 
independent of the power of 
the noise.  This allows networks 
of elements with
finite 
precision to take advantage of SR for diverse inputs.
To build upon the 
findings of Collins \etal, we consider networks of 
simpler model 
neurons, but these are allowed to have different dynamics.  
In 
particular, we 
examine noisy McCulloch-Pitts (\McP) neurons 
\citep{HerKroPal:1991} with a distribution of 
thresholds.  We construct 
heterogeneous networks that perform better than a 
homogeneous network 
with the same number of noisy \McP neurons and similar 
network 
architecture.

\section{Neural Network Model} \label{sec:netarch}

To investigate the effect of noise on signal transduction in networks of thresholding 
units, we consider a network of noisy McCulloch-Pitts (\McP) neurons. The \McP 
neuron is perhaps the simplest neural model, being only a simple thresholding unit. 
When the total input to a neuron (signal plus noise) exceeds its threshold, 
the neuron activates, firing an action potential or ``spike.''  Formally, the
activation state \(\activation{i}\) of neuron \(i\) in response to some 
stimulus \(\stimulus\) can be expressed as 
\begin{equation}
    \activation{i}\left(\stimulus\right) = \unitstep{\stimulus - \threshold}
    \label{eq:nonoiseneuron}
    \mathcomma
\end{equation}
where \( \threshold \) is the neuron's threshold and  \(\unitstepsymbol\) is the
Heaviside step function, defined as
\begin{equation}
    \unitstep{x} = \left \{ 
        \begin{array}{ll}
            1 & x > 0 \\
            0 & \mbox{otherwise}
        \end{array}
    \right . 
    \mathperiod
\end{equation}
We model the limited precision of neurons by including  noise as an 
intrinsic\footnote{Although we conceptually take the noise as intrinsic to the neuron,
the model we use is formally equivalent to a noise-free neuron subjected to a noisy 
stimulus.  } 
feature of the neurons, so that \eqn{eq:nonoiseneuron} becomes
\begin{equation}
    \activation{i}\left(\stimulus\right) = 
            \unitstep{\stimulus - \threshold + \noisedist}
    \label{eq:noisyneuron}
    \mathcomma
\end{equation}
where \( \noisedist \) is zero-mean, i.i.d.~(independent, identically distributed) 
Gaussian noise with variance \( \noisevar \).  
We assume the noise distributions to be identical for all the \McP neurons.

The 
network architecture is simple: an input layer of \( N  \) noisy 
\McP neurons is 
connected to a single linear output neuron.  Each synaptic weight is of 
identical 
and unitary strength, so the output neuron  calculates the sum
of the \( N \) input 
neuron activation states as its response \(\netresponse{N}(\stimulus) \):
\begin{equation}
    \netresponse{N} \left(\stimulus\right) = 
            \sum_{i=1}^{N} \activation{i}\left(\stimulus\right) 
    \label{eq:netdef}
    \mathperiod
\end{equation}   
Each input unit is presented the same analog signal, but 
with a different realization of 
the intrinsic neuronal noise.  

An important special case is when there is just a single input neuron (\(N = 1\)). 
Since the output neuron is a summing unit, its response is just the response 
of the single input neuron, 
\ie, \( \netresponse{N}(\stimulus) = \activation{1}(\stimulus) \).
In this chapter, we will use ``single neuron'' synonymously 
with ``network having only a single input neuron.''

\section{Network Response}\label{sec:throughput}

In this section, we consider the  response \( \netresponse{N} \) of the 
network. Due to the specific choices of neural model and network architecture made
in \sxn{sec:netarch}, the resulting neural networks are quite tractable mathematically.
A great deal of formal manipulation is thus possible, including exact calculations of
the expectation value and variance of the network response.

We initially focus on homogeneous networks with identical input neurons, including the 
special case of a single input neuron. 
The stimuli can be chosen without loss of generality so that \(\threshold=0\). The
results for \( \threshold \neq 0 \) can be recovered by a straightforward translation
along the \( \stimulus \)-axis.

The behavior for the standard, noise-free \McP neurons is trivial, with all 
input neurons synchronously firing or remaining quiescent.  
However, considerably more interesting behavior
is possible for noisy \McP neurons:
subthreshold signals have some chance of causing a neuron to fire, while 
suprathreshold signals have some chance of failing to cause the neuron to fire.

For a network with \(N\) input neurons
with the network architecture discussed above \sxn{sec:netarch},  
the response \( \netresponse{N} \) of the output neuron is just the number of input
neurons that fire. 
The probability \openprob{\stimulus}{\noisevar}\ of any neuron firing is 
\begin{equation}
    \openprob{\stimulus}{\noisevar} = \uppertailint{}{-\stimulus}{x}
    \mathcomma
    \label{eq:openprob}
\end{equation}
while the probability \closeprob{\stimulus}{\noisevar}\ for the neuron to remain quiescent is 
\begin{equation}
    \closeprob{\stimulus}{\noisevar} = \lowertailint{}{-\stimulus}{x}
    \mathperiod
    \label{eq:closeprob}
\end{equation}
Combining \eqns{eq:openprob} \andeqn{eq:closeprob} gives 
\(\closeprob{\stimulus}{\noisevar}+\openprob{\stimulus}{\noisevar}=1\), 
as expected.

\begin{figure}[tbp]
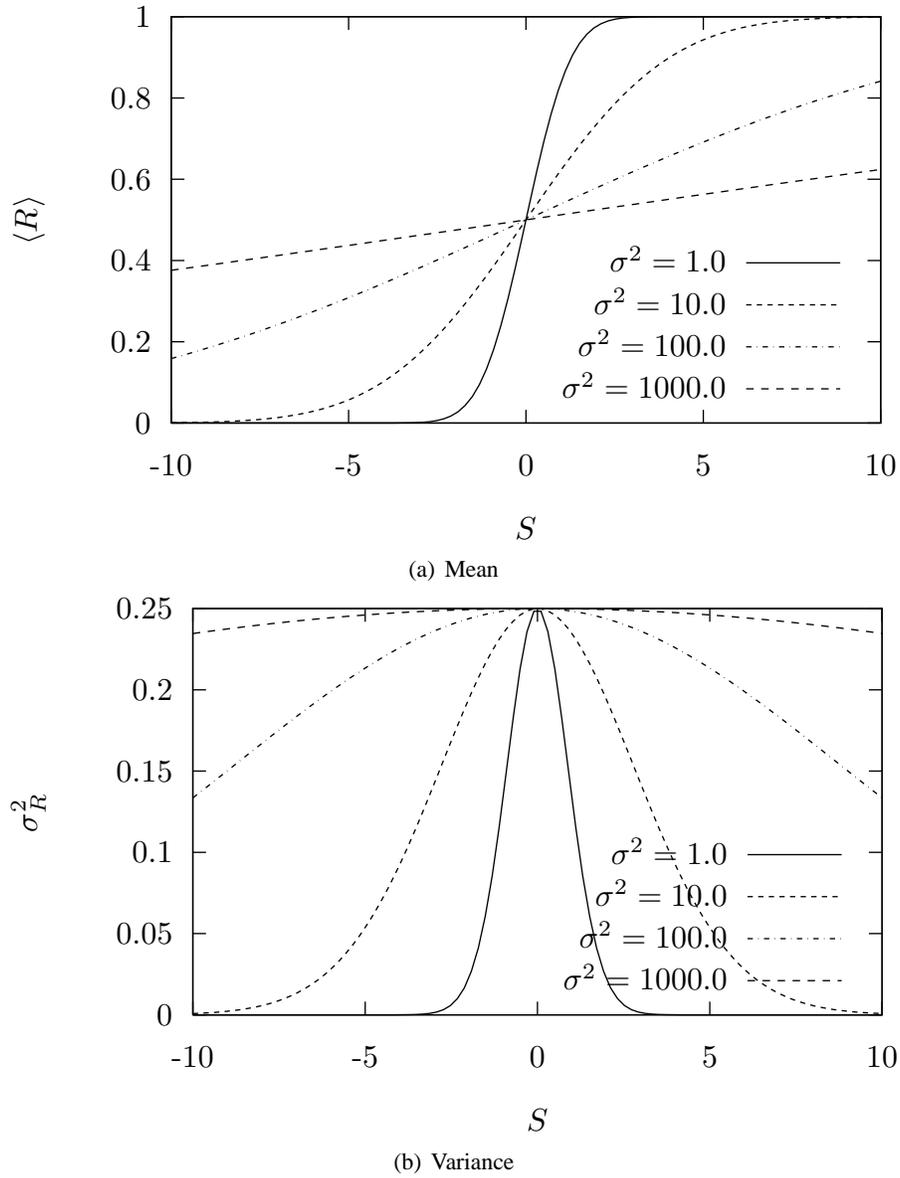

    \centering
	\subfigure[Mean]{
	    \includegraphics{channels0}
		\label{fig:netresponsemean}}
	\subfigure[Variance]{
	    \includegraphics{channels1}
		\label{fig:netresponsevariance}}
    \caption[Statistics of single neuron activation.] {
		Statistics of single neuron activation. 
		As the noise variance $\noisevar$ increases, (a) the mean activation state 
		\expectednetresponse{}{\stimulus}{\noisevar}
		takes longer to saturate to the extreme values, while (b) the 
		variance~\netresponsevar{}{\stimulus}{\noisevar}
		of the activation state increases with the noise variance.
	} \label{fig:singleneuronstatistics}
\end{figure}

Given \eqns{eq:openprob} \andeqn{eq:closeprob}, the expectation
value \(\expectednetresponse{}{\stimulus}{\noisevar}\) and 
variance \(\variance{Z}(\stimulus; \noisevar)\)
when \(N=1\) are
\begin{eqnarray}
    \expectednetresponse{}{\stimulus}{\noisevar} & = & \openprob{\stimulus}{\noisevar} \\
    \netresponsevar{}{\stimulus}{\noisevar} & = & 
				\openprob{\stimulus}{\noisevar} \closeprob{\stimulus}{\noisevar}
    \mathperiod
\end{eqnarray}
Since the noise is independent, the probability of different input neurons firing is also
independent and the expected value and variance of the output neuron activation 
are seen to be
\begin{eqnarray}
    \expectednetresponse{N}{\stimulus}{\noisevar} & = & 
			N \openprob{\stimulus}{\noisevar} \label{eq:expectednetresponseN} \\
    \netresponsevar{N}{\stimulus}{\noisevar} & = & 
			N \openprob{\stimulus}{\noisevar} \closeprob{\stimulus}{\noisevar}
    \mathperiod
\end{eqnarray}
The dependence of \( \expectednetresponse{}{\stimulus}{\noisevar} \)
and \(\netresponsevar{}{\stimulus}{\noisevar}\) on the stimulus \(\stimulus\) and
the noise variance \( \noisevar \) is shown in \fig{fig:singleneuronstatistics}.

\section{Decoding the Network Output}\label{sec:decoding}

In this section, we explore the ability of the neural circuit to serve as a signal transducer. 
We identify limits on the signal transduction capability by decoding the state of the output 
neuron to reproduce the input stimulus.  
Near the threshold value~\( \threshold \), this gives rise to linear decoding rules. 
The basic approach is similar to the  ``reverse reconstruction''  using linear filtering 
that has been applied with great effect to the analysis of a number of biological systems
\citep[see, \eg,][]{BiaRie:1992,BiaRieRuyWar:1991,RieWarRuyBia:1997,%
PraGabBra:2000,TheRodStuClaMil:1996}.

\begin{figure}[p]
    \centering
    \includegraphics{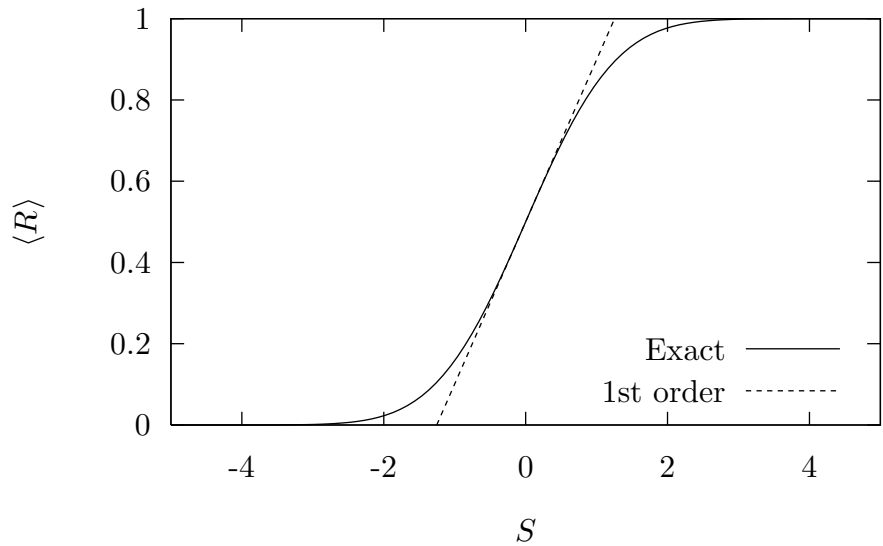}
    \caption[First order approximation of the expected activation.]{First order approximation 
    of the expected  activation of a single neuron. Near the threshold  ($\threshold = 0$), 
    the expected activation is nearly
    linear. Further from the threshold, the activation saturates at either zero or one and 
	diverges
    from the linear approximation. The values shown here are based on noise variance 
    $\noisevar=1$.}
    \label{fig:firstordermeannetresponse}
\end{figure}

We expand the expected output \( \netresponse{N} \) to first order near the
threshold (\ie, $\stimulus \approaches 0$), giving 
\begin{equation}
    \expectednetresponse{N}{\stimulus}{\noisevar} = 
		\frac{N}{2} + \frac{N}{\sqrt{2\pi\noisevar}}\stimulus + 						
					O\left(\stimulus^{2}\right)
    \mathperiod
    \label{eq:netresponseexpansion}
\end{equation}
An example of the linear approximation is shown in \fig{fig:firstordermeannetresponse}.

\begin{figure}[p]
    \centering
    \includegraphics{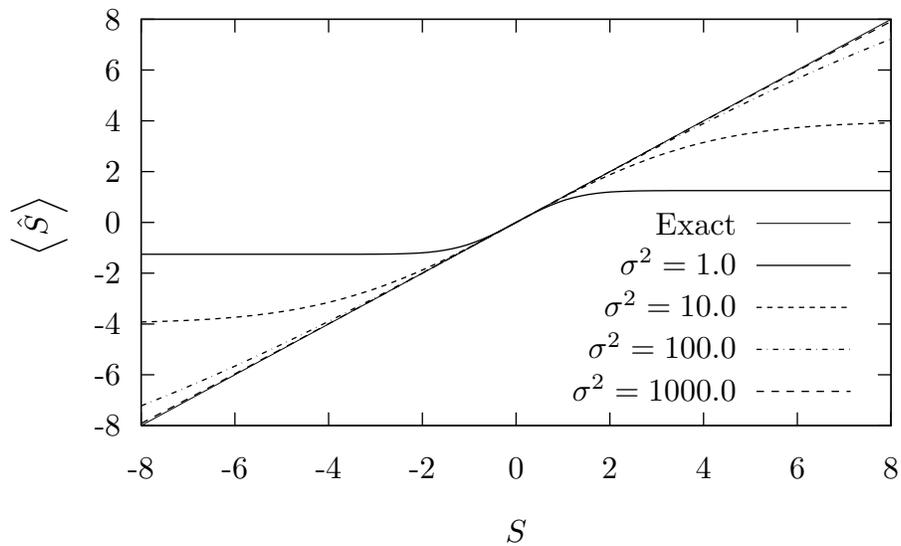}
    \caption[Expectation value of the decoded stimulus.]{Expectation value of the 
    stimulus decoded from the output of a single neuron. 
    As the noise variance $\noisevar$ increases, the expectation 
    value of the decoded stimulus approximates the true value of the stimulus 
	over an interval of increasing width. }
    \label{fig:decodingstim}
\end{figure}

Dropping the higher order terms and inverting \eqn{eq:netresponseexpansion} gives a 
linear decoding rule of the form
\begin{equation}
    \decodedstim{N} = \sqrt{2\pi\noisevar}\left(\frac{\netresponse{N}}{N} - \half \right) 
	\mathcomma
    \label{eq:decodingrule}
\end{equation}
where \(\decodedstim{N}\) is the estimate of the input stimulus. 
Combining \eqns{eq:expectednetresponseN} \andeqn{eq:decodingrule}, we can show that
\begin{equation}
    \expecteddecodedstim{N}{\stimulus}{\noisevar} = 
                \sqrt{2\pi\noisevar}\left(\openprob{\stimulus}{\noisevar} - \half \right) 
    \mathperiod
    \label{eq:meandecodedsignal}
\end{equation}
The expected value of 
\(\decodedstim{N}\) is thus seen to be independent of \(N\); for notational simplicity, we drop
the subscript and write \expectedstimestnoargs.
Examples of  \( \expectedstimestnoargs \) are shown in \fig{fig:decodingstim} for several 
values of the variance \( \noisevar \). Note that, as the noise variance increases, the 
expected value of the estimated stimulus closely matches the actual stimulus over 
a broader range.

\begin{figure}[p]
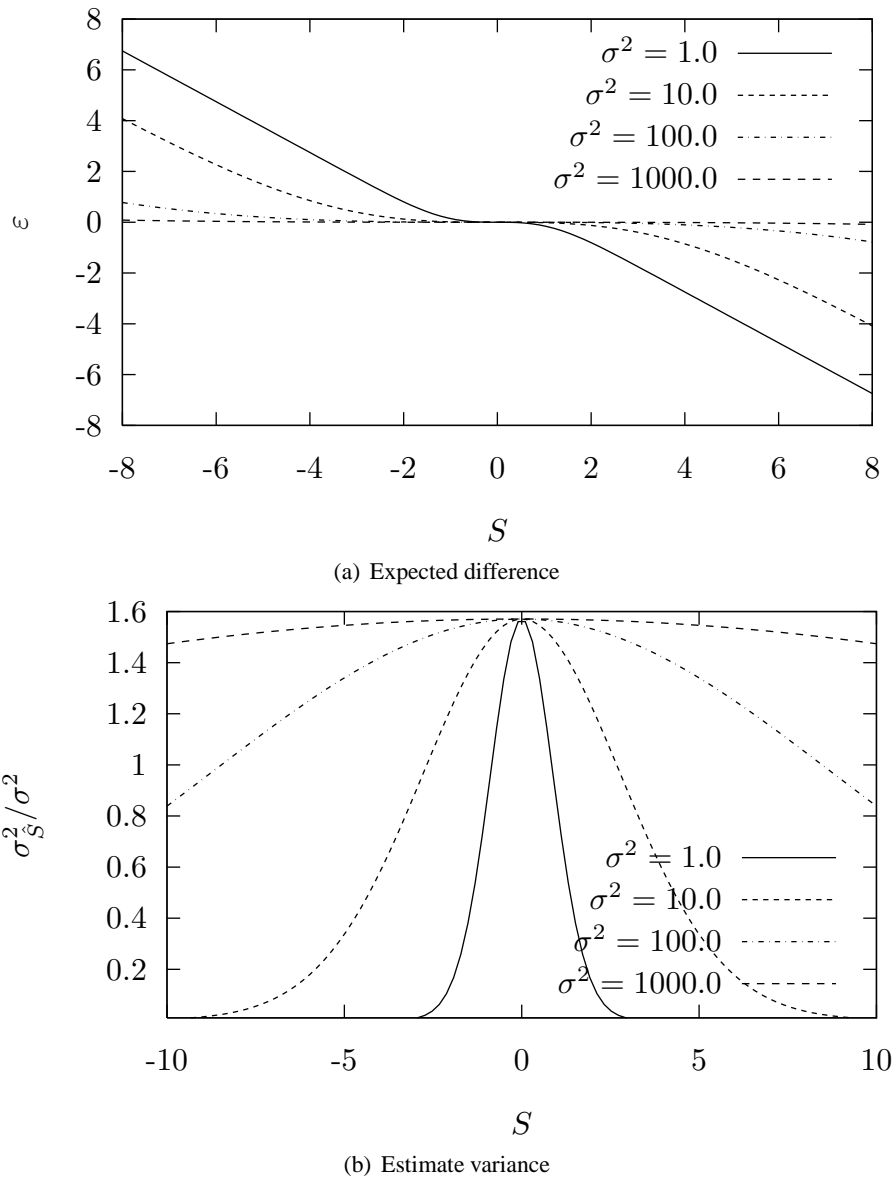

    \centering
	\subfigure[Expected difference]{
	    \includegraphics{channels4}
	    \label{fig:expecteddiffstim}
	}
	\subfigure[Estimate variance]{
	    \includegraphics{channels5}
	    \label{fig:decodingstimvariance}
	}
    \caption[Components of the decoding error.]
    {(a) Expected difference between the  stimulus and the value decoded 
    from the single-neuron response. The
    decoded value systematically diverges from the true value as the input gets
    farther from the threshold value at zero. (b) Variance of the 
    decoded stimulus values. Again, the variances
    shown here are based on decoding the single-neuron response. The variance of 
    the neuronal noise has been used to scale the variances of the estimates
    into a uniform range.}
	\label{fig:decodingerrcomponents}
\end{figure}

We must also consider the uncertainty of the value decoded from the network response. 
This leads to a total decoding error \decodingerror{N} with the form
\begin{eqnarray}
    \totalvariance{N}{\stimulus}{\noisevar} & = & 
    			\expectation{\left( \decodedstim{N} - \stimulus \right)^{2}} \nonumber \\
            & = & \expectedsquaredifference{\stimulus}{\noisevar} +
 					\decodingvariance{N}{\stimulus}{\noisevar} 
    \mathcomma
    \label{eq:decodingerror}
\end{eqnarray}
where
\begin{eqnarray}
    \expecteddifference{\stimulus}{\noisevar} & = & \expectedstimest{\stimulus}{\noisevar} - \stimulus
    \label{eq:expecteddifference} \\
    \decodingvariance{N}{\stimulus}{\noisevar} & = & 
        \expectation{\left( \decodedstim{N} 
                - \expectedstimest{\stimulus}{\noisevar} \right) ^{2} } \nonumber \\
    & = & \frac{2\pi\noisevar}{N}
                \openprob{\stimulus}{\noisevar}\closeprob{\stimulus}{\noisevar} 
    \mathperiod
    \label{eq:decodingvariance}    
\end{eqnarray}
The expected difference \( \expecteddifference{\stimulus}{\noisevar} \) and the decoding variance 
\decodingvariance{N}{\stimulus}{\noisevar} are shown in \figs{fig:decodingerrcomponents}.

\section{The Role of Noise}\label{sec:noise}

The noisy nature of the neurons has a striking and counter-intuitive effect on the
properties of the activation state: increasing noise improves
signal transmission, as seen 
in \figs{fig:decodingstim} \andfig{fig:expecteddiffstim}. This effect is analogous 
to the stochastic
resonance effect \citep{GamHanJunMar:1998}. SR can be informally understood as
the noise sometimes driving a nominally subthreshold signal to cross the threshold 
and producing
a current. Signals close to the threshold will more frequently cross the threshold, giving
a stronger response than signals far from the threshold.

There are several properties of the activation probabilities that we 
can derive from \eqns{eq:openprob} \andeqn{eq:closeprob} and that we will find useful
for understanding the role of noise in the neural behavior. First, there is a scaling 
property with the form
\begin{equation}
    \openprob{\stimulus}{\noisevar}  =  \openprob{\scaledstim}{\scalednoisevar} 
    \mathcomma
    \label{eq:openprobscaling}
\end{equation}
where \(\scalefactor > 0\).
Second, there is a reflection property with the form
\begin{equation}
    \openprob{\stimulus}{\noisevar} = \closeprob{-\stimulus}{\noisevar}
    \mathperiod
    \label{eq:openprobreflection} 
\end{equation}
As the activation probabilities are at the core of essentially all the equations
in this chapter, the scaling and reflection properties will be broadly useful to us.

\begin{figure}[p]
    \centering
	\subfigure[Mean] {
	    \includegraphics{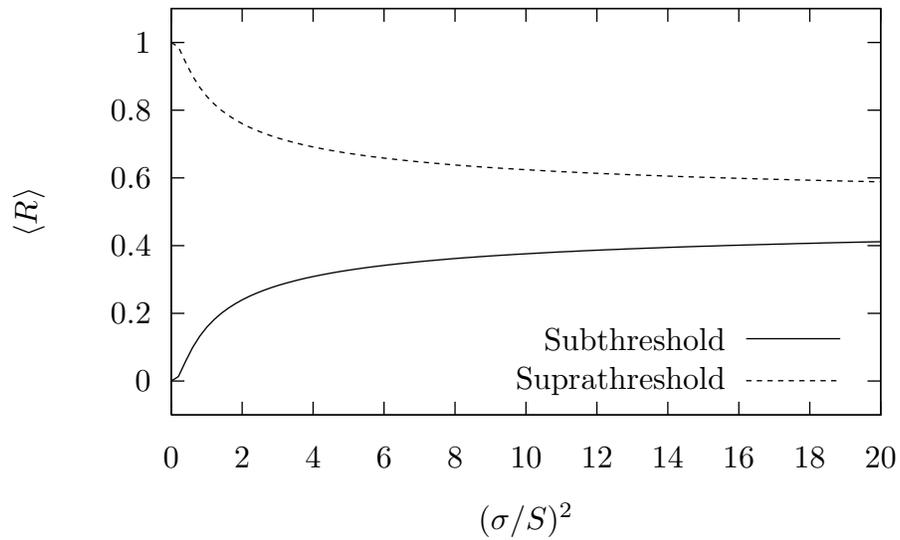}
		\label{fig:meannetresponsesig}
	}
	\subfigure[Variance]{    
		\includegraphics{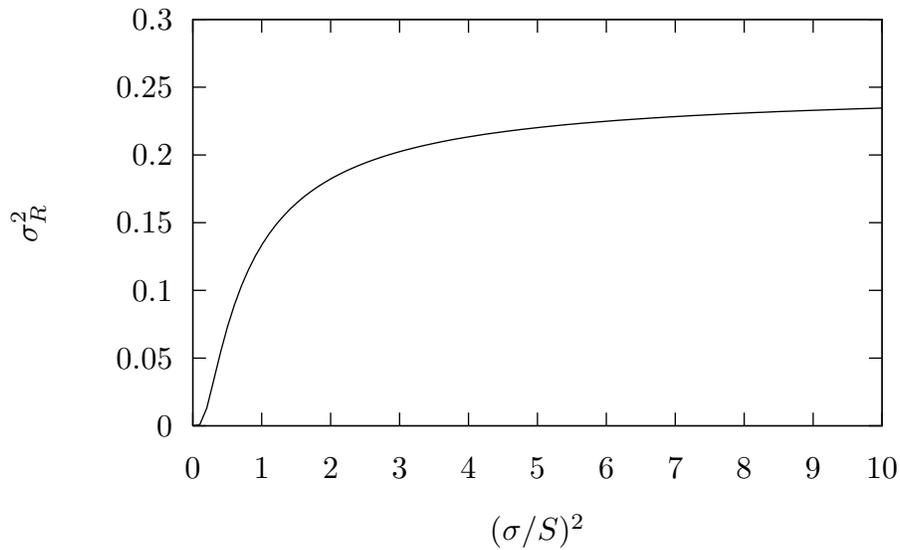}
    	\label{fig:netresponsevarsig}
	}    
	\caption[Noise dependence of activation statistics.]
    {(a) Noise dependence of single-neuron mean activation. 
    For large values of $(\noisestdev/\stimulus)^{2}$, the expected activation state 
	asymptotically approaches $1/2$.
	(b) Noise dependence of single-neuron activation variance. 
    For large 
    values of $(\sigma/V)^{2}$, the variance asymptotically approaches $1/4$.}
	\label{fig:netresponsestatisticssig}
\end{figure}

The scaling and reflection properties of the activation probabilities
can be used to derive similar properties of the neural responses. The statistics of the
neural responses obey the relations
\begin{eqnarray}
    \expectednetresponse{N}{\stimulus}{\noisevar} & = & 1 -
 				\expectednetresponse{N}{-\stimulus}{\noisevar} 
    \label{eq:currentreflect}\\
    \expectednetresponse{N}{\stimulus}{\noisevar} & = & 
				\expectednetresponse{N}{\scaledstim}{\scalednoisevar}
    \label{eq:currentscale}\\
    \netresponsevar{N}{\stimulus}{\noisevar} & = & \netresponsevar{N}{-\stimulus}{\noisevar} 
    \label{eq:currentvarreflect}\\
    \netresponsevar{N}{\stimulus}{\noisevar} & = & 
				\netresponsevar{N}{\scaledstim}{\scalednoisevar}
    \label{eq:currentvarscale}
    \mathcomma
\end{eqnarray}
where \(\scalefactor > 0\). Important corollaries of these relations are that 
\(\expectednetresponse{N}{\stimulus}{\noisevar} = 
\expectednetresponse{N}{-1}{(\sigma/V)^{2}}\) for all
\(V <0\), 
\(\expectednetresponse{N}{\stimulus}{\noisevar} = \expectednetresponse{N}{1}{(\sigma/V)^{2}}\) 
for all
\(V >0\), 
and \(\netresponsevar{N}{\stimulus}{\noisevar} = \netresponsevar{N}{1}{(\sigma/V)^{2}}\) 
for all \(V \neq 0\). It is thus necessary to consider only one 
subthreshold stimulus and one suprathreshold stimulus
in order to understand the impact of noise 
on the neural responses; see \fig{fig:netresponsestatisticssig}.

\begin{figure}[p]
    \centering
	\subfigure[Estimated stimulus]{    
		\includegraphics{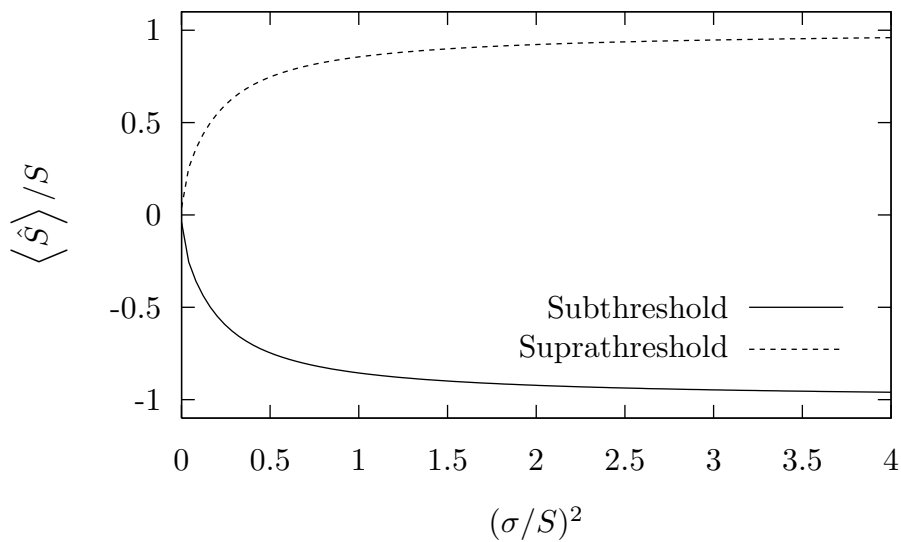}
    	\label{fig:decodingsig}
	}
	\subfigure[Expected difference]{ 
		\includegraphics{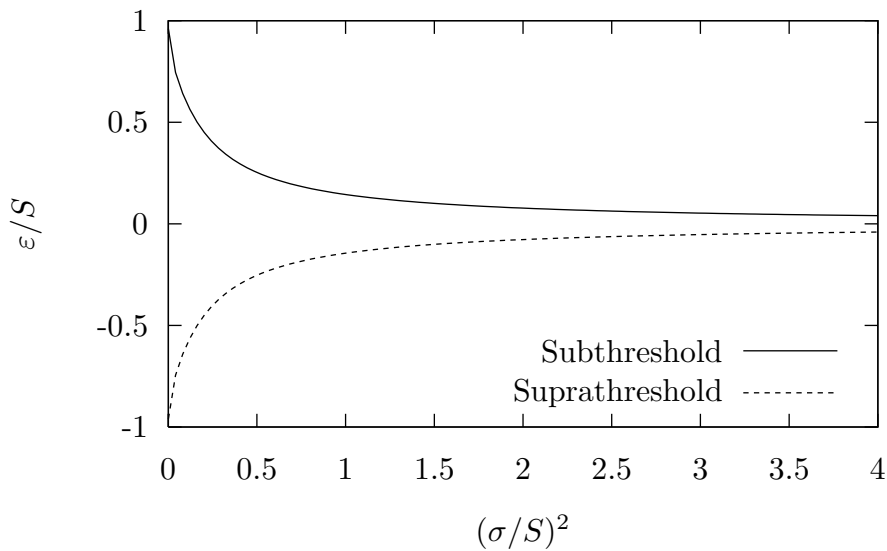}
	    \label{fig:expecteddiffsig}
	}
    \caption[Noise dependence of decoding quality.]
    {(a) Noise dependence of the single-neurons estimated stimulus. 
	For large values of $(\noisestdev/\stimulus)^{2}$, the estimates for the 
    subthreshold and suprathreshold signals asymptotically
    approach $-1$ and $+1$, respectively.
    (b) Noise dependence of the expected difference.
	For large values of $(\noisestdev/\stimulus)^{2}$, the expected differences 
    asymptotically approach 0 for 
    both the subthreshold and suprathreshold signals.}
	\label{fig:decodingnoisedependence}
\end{figure}

Similarly, properties of the statistics for the estimated input \(\decodedstim{}\) can be 
derived, giving
\begin{eqnarray}
    \estimatedinput{-\stimulus}{\noisevar} & = &  - \estimatedinput{\stimulus}{\noisevar} 
    \label{eq:estreflect}\\
    \estimatedinput{\scaledstim}{\scalednoisevar} & = & \scalefactor 
	\estimatedinput{\stimulus}{\noisevar}
    \label{eq:estscale}\\
    \expecteddifference{-\stimulus}{\noisevar} & = &  - \expecteddifference{\stimulus}{\noisevar} 
    \label{eq:estdiffreflect}\\
    \expecteddifference{\scaledstim}{\scalednoisevar} & = & 
			\scalefactor\expecteddifference{\stimulus}{\noisevar} 
    \label{eq:estdiffscale}\\
    \netresponsevar{N}{-\stimulus}{\noisevar} & = & \netresponsevar{N}{\stimulus}{\noisevar} 
    \label{eq:estvarreflect}\\
    \netresponsevar{N}{\scaledstim}{\scalednoisevar} & = &                                               
			\scalefactor^{2}\netresponsevar{N}{\stimulus}{\noisevar}
    \label{eq:estvarscale}
    \mathcomma
\end{eqnarray}
where \(\scalefactor > 0\). \Eqns{eq:estreflect} \througheqn{eq:estdiffscale} imply that
\(\estimatedinput{\stimulus}{\noisevar} =  
\stimulus\estimatedinput{1}{(\noisestdev/\stimulus)^{2}}\) and
\(\expecteddifference{\stimulus}{\noisevar} = 
\stimulus\expecteddifference{1}{(\noisestdev/\stimulus)^{2}}\) 
for all \(\stimulus \neq 0\).
Again, we can focus on one subthreshold stimulus and one suprathreshold
stimulus to understand the impact of noise (see \fig{fig:decodingnoisedependence}) 
for the behavior in the two cases, and use straightforward transformation to obtain the
exact results for other stimuli. 

\begin{figure}[p]
    \centering
    \includegraphics{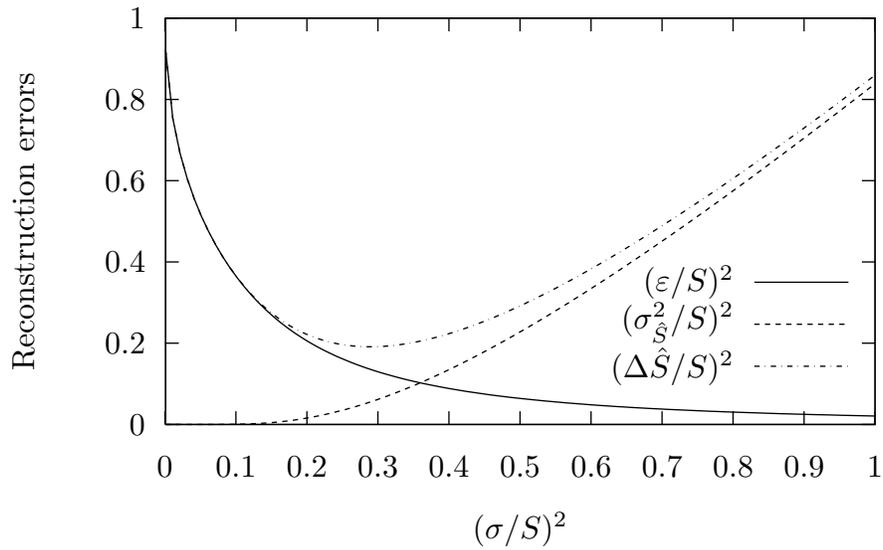}
    \caption[Comparison of decoding error sources.]
    {Comparison of decoding error sources. The values shown here are calculated
    from the response of a single neuron. The minimum in $\decodingerror{}^{2}$
    occurs for a nonzero noise variance of the signal, as with stochastic resonance.}
    \label{fig:decodingerrorcompare}
\end{figure}

Further, \eqn{eq:decodingerror} and \eqns{eq:estdiffreflect} \througheqn{eq:estvarscale}
imply
\(\expectedsquaredifference{\stimulus}{\noisevar} = 
\stimulus^{2}\expectedsquaredifference{1}{(\noisestdev/\stimulus)^{2}} \),
\(\netresponsevar{N}{\stimulus}{\noisevar} = 
\stimulus^{2}\netresponsevar{N}{1}{(\noisestdev/\stimulus)^{2}}\), and 
\(\totalvariance{N}{\stimulus}{\noisevar} = 
\stimulus^{2}\totalvariance{N}{1}{(\noisestdev/\stimulus)}\) 
for all \(\stimulus \neq 0\). Thus, the noise dependence of these latter error sources can be 
understood with a single stimulus; see \fig{fig:decodingerrorcompare}. Note that
the total error \(\totalvariance{N}{1}{(\noisestdev/\stimulus)^{2}}\) 
has its minimum for a nonzero value 
of the noise variance, analogous to the stochastic 
resonance effect; see \fig{fig:decodingerrorcompare}.

\section{Networks of Heterogeneous Neurons}\label{sec:populations}

Thus far, we have focused on single neurons and networks of identical neurons. The
effect of multiple neurons has generally been simple, either having no 
effect on---\expectedstimestnoargs, \expecteddiff---or just 
rescaling---\expectednetresponsenoargs{N}, \netresponsevarnoargs{N}, 
\decodingvariancenoargs{N}---the single-channel values. 

\begin{figure}[p]
    \centering
    \includegraphics{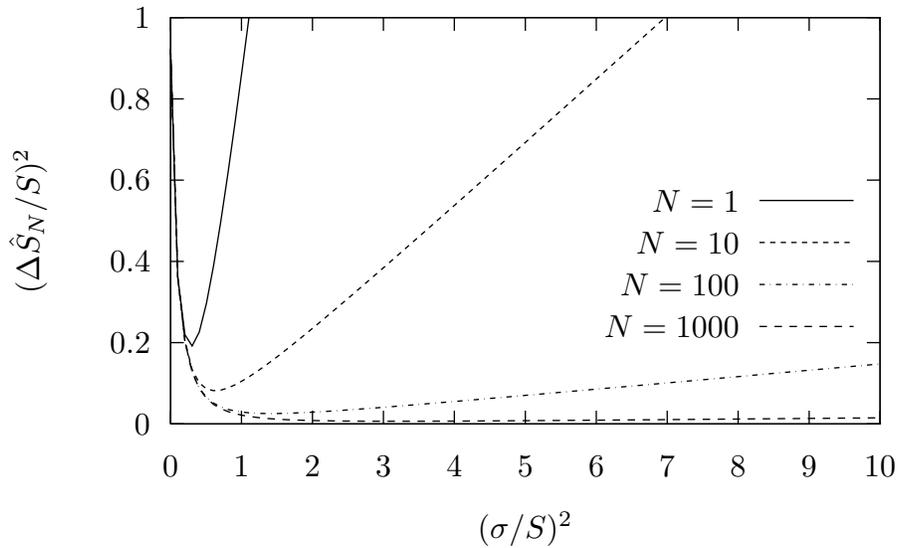}
    \caption[Effect of the number of neurons on the decoding error.]
    {Effect of the number of neurons on the decoding error. As $N$ becomes large, the
    error curve flattens out, indicating a broad range of noise values that all give similar 
    accuracy in the decoding process.}
    \label{fig:decodingerrorN}
\end{figure}

A significant exception to this general trend is found in \(\decodingerror{N}^{2}\). 
In \fig{fig:decodingerrorN}, we show how \(\decodingerror{N}^{2}\) varies with \(N\). 
The error curve flattens out into a broad range of similar values, so that the presence of
noise enhances signal transduction without requiring a precise relation 
between \(\stimulus\) and \noisevar\ seen for smaller values of \(N\). 
This effect is essentially the ``stochastic resonance 
without tuning'' first reported by \citet{ColChoImh:1995}.

Informally stated, SR without tuning allows for a wider range of potentials 
to be accurately decoded from the channel states for any particular value of 
the noise variance. To make this notion of ``wider range'' precise, we again focus our 
attention on the expected response of the neurons (see 
\fig{fig:firstordermeannetresponse}). The expected neural response 
\( \expectednetresponsenoargs{N} \) 
saturates to zero or one when \( \stimulus \) is far from the neuronal threshold. The
width \( \responsewidth \) of the intermediate range can be defined, for example, by
taking the boundaries of this range to be the points where the first order approximation
reaches the saturation values of zero and one. The width in this case becomes 
\( \responsewidth = \roottwopivar{} \). 

Other definitions for the response width 
are, of course, possible, but we still should observe that the width is 
proportional to \(\noisestdev\), since the activation probability 
depends only on the ratio of \(\stimulus\) and \(\noisestdev\) (\eqn{eq:openprob}). 
The same width is 
found for multiple identical input neurons, because the output neuron response is 
proportional to the single neurons response, without broadening the curve in 
\fig{fig:firstordermeannetresponse}.

The response width can thus be increased by increasing the noise variance \noisevar. As seen in \figs{fig:decodingerrorcompare} \andfig{fig:decodingerrorN}, such an
increase ultimately leads to a growth in the decoding error \(\decodingerror{N}^{2}\). In the asymptotic limit as \noisevar\ becomes large, \(\decodingerror{N}^{2}\) is dominated by
\variance{\decodedstim{N}} and
we have the asymptotic behavior 
\begin{equation}
    \totalvariance{N}{\stimulus}{\noisevar} = O\left( \frac{\noisevar}{N} \right) 
    \mathcomma
    \label{eq:asymptoticerror}
\end{equation}
based on \eqn{eq:decodingvariance}.
The growth in \(\decodingerror{N}^{2}\) with increasing \( \noisevar \) thus can be overcome by
further increasing the number of neurons in the input layer. Therefore, the response 
width \(\responsewidth\) is effectively constrained by the number of neurons \(N\), 
with \(W=O(\sqrt{N})\) for large \(N\). 

An arbitrary response width can be produced by assembling enough neurons. However, this approach 
is inefficient, and greater width increases can be achieved with the same number of neurons. 
Consider instead dividing up the total width into \(M\) subranges. These subranges can each be 
independently covered by a subpopulation of \(N\) neurons; all neurons within a subpopulation
are identical to one another, while neurons from different subpopulations differ only in
their thresholds.
The width of each subrange is \(O(\sqrt{N})\), but the total width is \(O(M\sqrt{N})\). Thus, the 
total response width can increase 
more rapidly as additional subpopulations of neurons are added.
Conceptually, multiple thresholds are a way to provide a wide range of accurate responses, with multiple neurons in each subpopulation providing independence from any need to ``tune'' the 
noise variance to a particular value.

To describe the behavior of channels with different thresholds, much 
of the preceding analysis can be directly applied  by translating the functions 
along the potential axis to obtain the desired threshold. 
However, 
system behavior was previously explored near the threshold
value, but  heterogeneous populations of neurons have multiple thresholds. Nonetheless, 
we can produce a comparable system by simply assessing system behavior
near the center of the total response width.

To facilitate a clean comparison, 
we set the thresholds in the heterogeneous populations so that a linear decoding rule can be 
readily
produced. A simple approach that achieves this is to space the thresholds of the subpopulations 
by \(2\responsewidth\), with all neurons being otherwise equal. 
The subpopulations with lower thresholds provide an upward shift in the expected 
number of active neurons for higher threshold subpopulations, such that the different 
subpopulations are all approximated to first order by the same line. Thus, the expected 
total number of active neurons leads to a 
linear decoding rule by expanding to first order and inverting, as was done 
earlier
for homogeneous populations. Note that this construction requires 
no additional assumptions about how the neural responses are to be interpreted, nor does it 
require alterations to the network architecture. 

To illustrate the effect of multiple thresholds, we begin by investigating the response 
of a homogeneous baseline to a stimulus \( \stimulus \). The baseline
network consists of  \(M=1\) populations of \(N=1000\) neurons with \(\threshold=0\) and
variance \(\variance{}=1\). Using the definition above, the response width 
is \(W=\sqrt{2\pi}\). We then consider two cases, homogeneous and heterogeneous, in each of 
which we increase the response width by doubling the number of neurons while maintaining 
similar error expectations for the decoded stimuli. 

\begin{figure}[p]
    \centering
	\subfigure[Decoded stimulus]{
	    \includegraphics{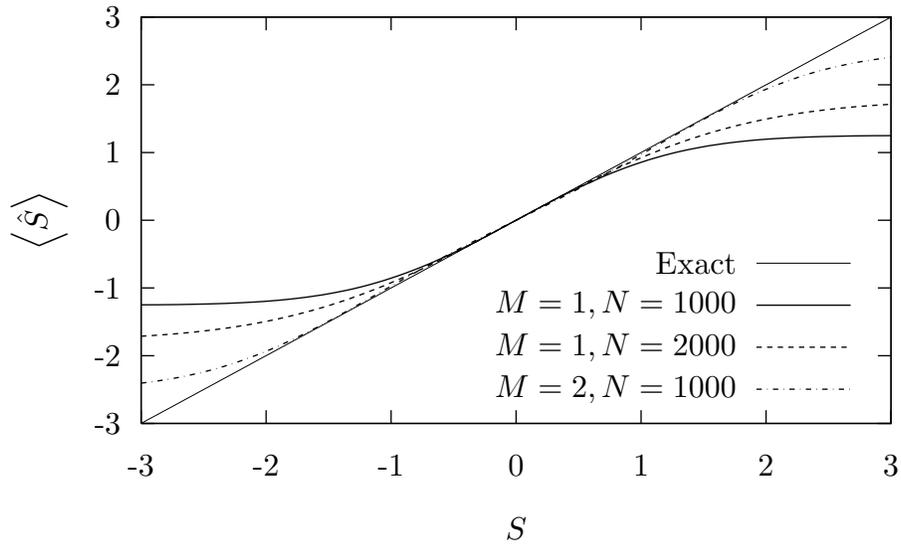}
	    \label{fig:heterogeneousdecoding}
	}
	\subfigure[Decoding error]{
	    \includegraphics{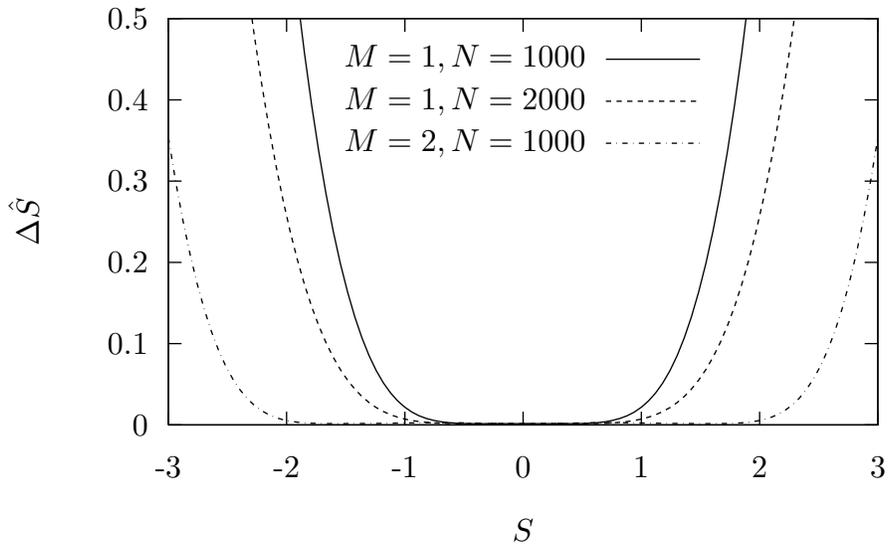}
	    \label{fig:heterogeneousdecodingerror}
	}
    \caption[Decoding in homogeneous and heterogeneous networks.]
    {(a) Expectation value of the decoded output in homogeneous and heterogeneous
    networks. The response of the heterogeneous neural network ($M=2, N=1000$)
    can be accurately decoded  over a broader range than the responses of the 
	baseline ($M=1, N=1000$) and homogeneous ($M=1, N=2000$) networks.
    (b) Total decoding error for homogeneous and heterogeneous
    networks. The heterogeneous neural network
    has a broader basin of low error values than  the baseline 
    and homogeneous networks.}
	\label{fig:heterogeneousdecodingcomparison}
\end{figure}

In the homogeneous case, we have a single population (\(M=1\)) with \(N=2000\) neurons. 
Doubling the number of neurons allows us to double the variance to \(\variance{}=2\) 
with similar expected errors outside the response width. Thus, we observe an extended 
range, relative to the baseline case, in which we can reconstruct the stimulus  
from the network output (\fig{fig:heterogeneousdecodingcomparison}). 

In the heterogeneous case, we instead construct two subpopulations (\(M=2\)) with \(N=1000\) 
neurons. We leave the variance unchanged at \(\variance{}=1\). One of the subpopulations is 
modified so that the thresholds lie at \(+W/2=\sqrt{\pi/2}\), while the other is modified so that 
the  thresholds lie at \(-W/2=-\sqrt{\pi/2}\). The resulting neural network has a broad range in 
which we can reconstruct the stimulus from the network response, markedly greater than the 
baseline and homogeneous cases (\fig{fig:heterogeneousdecodingcomparison}).

\section{Conclusion}\label{sec:discussion}

We have constructed networks of heterogeneous \McP neurons  
that
outperform similar 
networks of homogeneous \McP neurons.  The   network architectures 
are identical, with the 
only difference being the distribution of 
neuronal thresholds.  The heterogeneous networks 
are
sensitive to a wider range of signals than the homogeneous networks.  
Such networks are easily implemented, and could serve as simple models of many 
diverse natural and artificial systems.

The superior scaling
properties of heterogeneous neuronal networks can have a profound metabolic impact; 
large numbers of neurons imply a large energetic investment, in terms of both 
cellular maintenance and neural activity. The action potentials 
generated in neurons can require a significant energetic 
cost \citep{LauRuyAnd:1998}, making the 
trade-off between reliably coding information and the metabolic costs potentially 
quite important. Thus, we expect neuronal heterogeneity to be evolutionarily 
favored, even for quite simple neural circuits.

Although we have used a specific model consisting of thresholding neurons with 
additive Gaussian noise,
we expect that the key result is 
more widely applicable. The demonstration of the advantage of neuronal heterogeneity
largely follows from two factors that are not specific to the model 
neurons. First, the distance of the input stimulus from the threshold is proportional 
to the standard deviation of the Gaussian noise, and, second, the total variance of 
the network response 
is proportional to the number of input neurons. Ultimately, the heterogeneous thresholds 
are favorable because the independently distributed noise provides a natural scale for 
the system.

\section*{Acknowledgments}

We would like to acknowledge support from the Portuguese Funda\c{c}\~ao para a Ci\^encia e 
a Tecnologia under Bolsa de Investiga\c{c}\~ao SFRH/BPD/9417/2002 and 
Plurianual CCM.
The writing of this chapter was supported in part by ARC systems research GmbH.

\bibliographystyle{plainnat}
\bibliography{novasci}

\end{document}

%% file: novascimacros.tex
\usepackage{xspace}

\newcommand{\etal}{\textit{et al.}}

\newcommand{\ie}{i.e.}
 
\newcommand{\eg}{e.g.}

\newcommand{\email}[1]{\texttt{#1}}

\newcommand{\half}{\frac{1}{2}}
\newcommand{\expectation}[1]{\ensuremath{\left \langle #1 \right \rangle}}
\newcommand{\approaches}{\rightarrow}

\newcommand{\puncspace}{\enspace}
\newcommand{\mathcomma}{\puncspace ,}
\newcommand{\mathperiod}{\puncspace .}

\newcommand{\unitstepsymbol}{\ensuremath{u}}
\newcommand{\unitstep}[1]{\ensuremath{\unitstepsymbol\left(#1\right)}}

\newcommand{\McP}{McP\xspace}

\newcommand{\stimvarargs}[2]{\ensuremath{\left (#1; #2 \right)}}
\newcommand{\withstimvarargs}[3]{\ensuremath{#1 \stimvarargs{#2}{#3}}}

\newcommand{\formattedintegral}[4]{\int_{#1}^{#2}#3\,d#4}

\newcommand{\stimulus}{\ensuremath{S}}
\newcommand{\threshold}{\ensuremath{\stimulus_{0}}}
\newcommand{\activation}[1]{\ensuremath{a_{#1}}}

\newcommand{\netresponsesymbol}{\ensuremath{R}}
\newcommand{\netresponse}[1]{\ensuremath{\netresponsesymbol_{#1}}}
\newcommand{\expectednetresponsenoargs}[1]{\expectation{\netresponse{#1}}}
\newcommand{\expectednetresponse}[3]
				{\withstimvarargs{\expectednetresponsenoargs{#1}}{#2}{#3}}
\newcommand{\netresponsevarnoargs}[1]{\ensuremath{\variance{\netresponse{#1}}}}
\newcommand{\netresponsevar}[3]{\withstimvarargs{\netresponsevarnoargs{#1}}{#2}{#3}}

\newcommand{\stimest}{\ensuremath{\hat{\stimulus}}}
\newcommand{\expectedstimestnoargs}{\expectation{\stimest}}
\newcommand{\expectedstimest}[2]
				{\withstimvarargs{\expectedstimestnoargs}{#1}{#2}}
\newcommand{\decodedstim}[1]{\ensuremath{\stimest_{#1}}}
\newcommand{\expecteddecodedstimnoargs}[1]{\expectation{\decodedstim{#1}}}
\newcommand{\expecteddecodedstim}[3]
				{\withstimvarargs{\expecteddecodedstimnoargs{#1}}{#2}{#3}}
\newcommand{\decodingerror}[1]{\ensuremath{\Delta\decodedstim{#1}}}
\newcommand{\decodingvariancenoargs}[1]{\variance{\decodedstim{#1}}}
\newcommand{\decodingvariance}[3]
				{\withstimvarargs{\decodingvariancenoargs{#1}}{#2}{#3}}
\newcommand{\expecteddiff}{\ensuremath{\varepsilon}}
\newcommand{\expecteddifference}[2]{\withstimvarargs{\expecteddiff}{#1}{#2}}
\newcommand{\expectedsquaredifference}[2]
				{\withstimvarargs{\expecteddiff^{2}}{#1}{#2}}
\newcommand{\estimatedinput}[2]{\expecteddecodedstim{}{#1}{#2}}

\newcommand{\totalvariance}[3]{\ensuremath{\decodingerror{#1}^{2}\left(#2; #3\right)}}

\newcommand{\scalefactor}{\ensuremath{\alpha}}
\newcommand{\scaledstim}{\scalefactor \stimulus}
\newcommand{\scalednoisevar}{\scalefactor^{2}\noisevar}

\newcommand{\responsewidth}{\ensuremath{W}}

\newcommand{\noisedist}{\ensuremath{\eta}}
\newcommand{\stdev}{\sigma}
\newcommand{\variance}[1]{\ensuremath{\stdev_{#1}^{2}}}
\newcommand{\twopivar}[1]{2\pi\variance{#1}}
\newcommand{\roottwopivar}[1]{\sqrt{\twopivar{#1}}}
\newcommand{\noisevar}{\variance{}}
\newcommand{\noisestdev}{\stdev}
\newcommand{\openprob}[2]{\ensuremath{p\left(#1; #2\right)}}
\newcommand{\closeprob}[2]{\ensuremath{q\left(#1; #2\right)}}
\newcommand{\uppertailint}[3]{\frac{1}{\roottwopivar{#1}}
            \formattedintegral{#2}{\infty}{\exp\left(-\frac{#3^{2}}{2\variance{#1}}\right)}{#3}}
\newcommand{\lowertailint}[3]{\frac{1}{\roottwopivar{#1}}
            \formattedintegral{-\infty}{#2}{\exp\left(-\frac{#3^{2}}{2\variance{#1}}\right)}{#3}}

%% file: xrefmacros.tex
\newcommand{\wraprefprepost}[3]{\wrapprepost{#1}{#2}{\ref{#3}}}
\newcommand{\formatrefplain}{\ref}
\newcommand{\formatrefparens}{\wraprefprepost{(}{)}}

\newcommand{\wrapprepost}[3]{{#1}{#3}{#2}}

\newcommand{\tagwithlabel}[2]{#1~#2}

\newcommand{\makelabeledcrossrefmacro}[4]
	{\newcommand{#3}{#1{#4}{#2}}}
\newcommand{\makecrossrefmaker}[3]
	{\newcommand{#1}{\makelabeledcrossrefmacro{#2}{#3}}}

\newcommand{\eqnrefformat}{\formatrefparens}
\newcommand{\eqnlabelbinding}{\tagwithlabel}
\newcommand{\eqnlabel}{eq.}
\newcommand{\Eqnlabel}{Eq.}
\newcommand{\eqnslabel}{eqs.}
\newcommand{\Eqnslabel}{Eqs.}

\newcommand{\eqnnum}{\eqnrefformat}
\makecrossrefmaker{\newlabeledeqnref}{\eqnlabelbinding}{\eqnnum}
\makecrossrefmaker{\newwordpluseqnref}{\tagwithlabel}{\eqnnum}

\newlabeledeqnref{\eqn}{\eqnlabel}
\newlabeledeqnref{\Eqn}{\Eqnlabel}
\newlabeledeqnref{\eqns}{\eqnslabel}
\newlabeledeqnref{\Eqns}{\Eqnslabel}

\newwordpluseqnref{\andeqn}{and}
\newwordpluseqnref{\througheqn}{through}

\newcommand{\figrefformat}{\formatrefplain}
\newcommand{\figlabelbinding}{\tagwithlabel}
\newcommand{\figlabel}{fig.}
\newcommand{\Figlabel}{Fig.}
\newcommand{\figslabel}{figs.}
\newcommand{\Figslabel}{Figs.}

\newcommand{\fignum}{\figrefformat}
\makecrossrefmaker{\newlabeledfigref}{\figlabelbinding}{\fignum}
\makecrossrefmaker{\newwordplusfigref}{\tagwithlabel}{\fignum}

\newlabeledfigref{\fig}{\figlabel}
\newlabeledfigref{\Fig}{\Figlabel}
\newlabeledfigref{\figs}{\figslabel}
\newlabeledfigref{\Figs}{\Figslabel}

\newwordplusfigref{\andfig}{and}
\newwordplusfigref{\throughfig}{through}

\newcommand{\sxnrefformat}{\formatrefplain}
\newcommand{\sxnlabelbinding}{\tagwithlabel}
\newcommand{\sxnlabel}{section}
\newcommand{\Sxnlabel}{Section}
\newcommand{\sxnslabel}{sections}
\newcommand{\Sxnslabel}{Sections}

\newcommand{\sxnnum}{\sxnrefformat}
\makecrossrefmaker{\newlabeledsxnref}{\sxnlabelbinding}{\sxnnum}
\makecrossrefmaker{\newwordplussxnref}{\tagwithlabel}{\sxnnum}

\newlabeledsxnref{\sxn}{\sxnlabel}
\newlabeledsxnref{\Sxn}{\Sxnlabel}
\newlabeledsxnref{\sxns}{\sxnslabel}
\newlabeledsxnref{\Sxns}{\Sxnslabel}
	
\newwordplussxnref{\andsxn}{and}
\newwordplussxnref{\throughsxn}{through}

\newcommand{\tblrefformat}{\formatrefplain}
\newcommand{\tbllabelbinding}{\tagwithlabel}
\newcommand{\tbllabel}{table}
\newcommand{\Tbllabel}{Table}
\newcommand{\tblslabel}{tables}
\newcommand{\Tblslabel}{Tables}

\newcommand{\tblnum}{\tblrefformat}
\makecrossrefmaker{\newlabeledtblref}{\tbllabelbinding}{\tblnum}
\makecrossrefmaker{\newwordplustblref}{\tagwithlabel}{\tblnum}

\newlabeledtblref{\tbl}{\tbllabel}
\newlabeledtblref{\Tbl}{\Tbllabel}
\newlabeledtblref{\tbls}{\tblslabel}
\newlabeledtblref{\Tbls}{\Tblslabel}
	
\newwordplustblref{\andtbl}{and}
\newwordplustblref{\throughtbl}{through}